\documentclass{INTERSPEECH2023}

\interspeechcameraready

\usepackage{graphicx}
\usepackage{subfigure}
\usepackage{hyperref}

\title{Exploration on HuBERT with Multiple Resolutions}
\name{Jiatong Shi$^1$, Yun Tang$^2$, Hirofumi Inaguma$^2$, Hongyu Gong$^2$, Juan Pino$^2$, Shinji Watanabe$^1$}
\address{
  $^1$Language Technologies Institute, Carnegie Mellon University \quad $^2$Meta AI}
\email{\{jiatongs, swatanab\}@cs.cmu.edu}

\usepackage[
backend=biber,
style=ieee,
citestyle=numeric-comp,
maxbibnames=3,
maxcitenames=3,
doi=false,isbn=false,url=false,eprint=false
]{biblatex}

\defbibheading{bibliography}[\refname]{}

\DeclareSourcemap{
	\maps[datatype=bibtex, overwrite=true]{
		\map{
			\step[fieldsource=booktitle,
			match=\regexp{.*Interspeech.*},
			replace={Proc. Interspeech}]
			\step[fieldsource=journal,
			match=\regexp{.*INTERSPEECH.*},
			replace={Proc. Interspeech}]
			\step[fieldsource=booktitle,
			match=\regexp{.*ICASSP.*},
			replace={Proc. ICASSP}]
			\step[fieldsource=booktitle,
			match=\regexp{.*icassp_inpress.*},
			replace={Proc. ICASSP (in press)}]
			\step[fieldsource=booktitle,
			match=\regexp{.*Acoustics,.*Speech.*and.*Signal.*Processing.*},
			replace={Proc. ICASSP}]
			\step[fieldsource=booktitle,
			match=\regexp{.*International.*Conference.*on.*Learning.*Representations.*},
			replace={Proc. ICLR}]
			\step[fieldsource=booktitle,
			match=\regexp{.*International.*Conference.*on.*Computational.*Linguistics.*},
			replace={Proc. COLING}]
			\step[fieldsource=booktitle,
			match=\regexp{.*SIGdial.*Meeting.*on.*Discourse.*and.*Dialogue.*},
			replace={Proc. SIGDIAL}]
			\step[fieldsource=booktitle,
			match=\regexp{.*International.*Conference.*on.*Machine.*Learning.*},
			replace={Proc. ICML}]
			\step[fieldsource=booktitle,
			match=\regexp{.*North.*American.*Chapter.*of.*the.*Association.*for.*Computational.*Linguistics:.*Human.*Language.*Technologies.*},
			replace={Proc. NAACL}]
			\step[fieldsource=booktitle,
			match=\regexp{.*Empirical.*Methods.*in.*Natural.*Language.*Processing.*},
			replace={Proc. EMNLP}]
			\step[fieldsource=booktitle,
			match=\regexp{.*Association.*for.*Computational.*Linguistics.*},
			replace={Proc. ACL}]
			\step[fieldsource=booktitle,
			match=\regexp{.*Automatic.*Speech.*Recognition.*and.*Understanding.*},
			replace={Proc. ASRU}]
			\step[fieldsource=booktitle,
			match=\regexp{.*Spoken.*Language.*Technology.*},
			replace={Proc. SLT}]
			\step[fieldsource=booktitle,
			match=\regexp{.*Speech.*Synthesis.*Workshop.*},
			replace={Proc. SSW}]
			\step[fieldsource=booktitle,
			match=\regexp{.*workshop.*on.*speech.*synthesis.*},
			replace={Proc. SSW}]
			\step[fieldsource=booktitle,
			match=\regexp{.*Advances.*in.*neural.*information.*processing.*},
			replace={Proc. NeurIPS}]
			\step[fieldsource=booktitle,
			match=\regexp{.*Advances.*in.*Neural.*Information.*Processing.*},
			replace={Proc. NeurIPS}]
			\step[fieldsource=booktitle,
			match=\regexp{.*Workshop.*on.* Applications.* of.* Signal.*Processing.*to.*Audio.*and.*Acoustics.*},
			replace={Proc. WASPAA}]
			\step[fieldsource=publisher,
			match=\regexp{.+},
			replace={{}}]
			\step[fieldsource=month,
			match=\regexp{.+},
			replace={{}}]
			\step[fieldsource=location,
			match=\regexp{.+},
			replace={{}}]
			\step[fieldsource=address,
			match=\regexp{.+},
			replace={{}}]
			\step[fieldsource=organization,
			match=\regexp{.+},
			replace={{}}]
		}
	}
}
\begin{document}

\maketitle
\vspace{-50pt}
 
\begin{abstract}
Hidden-unit BERT (HuBERT) is a widely-used self-supervised learning (SSL) model in speech processing. However, we argue that its fixed 20ms resolution for hidden representations would not be optimal for various speech-processing tasks since their attributes (e.g., speaker characteristics and semantics) are based on different time scales. To address this limitation, we propose utilizing HuBERT representations at multiple resolutions for downstream tasks. We explore two approaches, namely the parallel and hierarchical approaches, for integrating HuBERT features with different resolutions. Through experiments, we demonstrate that HuBERT with multiple resolutions outperforms the original model. This highlights the potential of utilizing multiple resolutions in SSL models like HuBERT to capture diverse information from speech signals.
\end{abstract}

\section{Introduction}
\label{sec: intro}

In recent years, self-supervised learning (SSL) models for speech processing have demonstrated impressive performance on a range of tasks \cite{mohamed2022self}. These models can leverage unlabeled speech data to learn general-purpose knowledge, rather than relying solely on supervised training with paired labels. As a result, speech SSLs have emerged as a powerful tool for speech processing, offering a promising alternative to traditional supervised learning approaches.

HuBERT \cite{hsu2021hubert} is one of the most prominent speech self-supervised learning (SSL) models, according to the SUPERB benchmark \cite{yang21c_interspeech, tsai2022superb, feng2023superb, }. During training, HuBERT employs an offline clustering step to generate pseudo labels and uses a masked language model (MLM) objective. Like many speech processing systems, HuBERT begins by converting the raw waveform into hidden states using convolutional (conv.) layers, resulting in a fixed 20ms resolution for its representation. 

HuBERT can be used as a feature extractor or directly fine-tuned as an encoder. In the feature extraction approach, the pre-trained model is used to extract high-level features from speech signals, which are then fed into a downstream task-specific model such as a classifier or a regression model. This approach reduces the computational cost during training \cite{chang2021exploration, berrebbi22_interspeech}. On the other hand, fine-tuning the pre-trained HuBERT model as an encoder is a popular approach, which further improves the performance at the cost of training massive encoder parameters. In this approach, the pre-trained model is further trained on the downstream task data, either by updating all the model parameters or just the last few layers. 

Although the HuBERT representation has demonstrated strong performance, the empirical selection of a 20ms resolution raises concerns regarding its optimality for diverse speech-related tasks.\footnote{It's worth noting that this choice of resolution is derived from an ASR conversion involving downsampling, which is specific to that particular task.} In contrast, the literature also suggests that modeling speech at multiple resolutions is preferable for speech recognition \cite{mallidi2016novel, mallidi2018practical, hermansky2013multistream, han2021multistream, luo2021multi, li2019multi, kong2021multi, andrusenko2022uconv, kimsqueezeformer}, speaker verification \cite{yao2020multi, gao2022unet, burchi2021efficient}, speech enhancement \cite{zhang2019research, zhao2021unet++}, and voice conversion \cite{li2022unet}. Two mainstream approaches have emerged: one that focuses on parallel processing \cite{mallidi2016novel, mallidi2018practical, hermansky2013multistream, han2021multistream, luo2021multi, li2019multi, kong2021multi, yao2020multi}, and the other that utilizes hierarchical frameworks such as U-net \cite{gao2022unet, zhao2021unet++, li2022unet, xiang2021nested, xian2020multi, liu2020cp, xiang2021convolutional}.

The parallel paradigm is based on observations of multiple parallel processing streams in the human speech cognitive system \cite{mallidi2016novel, mallidi2018practical}. To formalize multi-stream signals, a common method is to use parallel encoders that consider multi-resolution signals. For example, \cite{li2019multi} employs two encoders based on recurrent neural network (RNN) and convolution neural network (CNN)-RNN, respectively. Both encoders use the same input features, while the second applies CNN to transform features into a different temporal resolution.

The second hierarchical approach, in contrast, serializes the aggregation of multi-resolution information. An example of this approach is the U-net-like architecture, which is based on an encoder-decoder structure \cite{gao2022unet, zhao2021unet++, li2022unet, andrusenko2022uconv, kimsqueezeformer}. The encoder processes high-resolution features initially and downsamples them to prioritize low-resolution features. Conversely, the decoder starts from low-resolution features and upsamples them to refine information in high resolution. To ensure stability, corresponding blocks with the same resolution in the encoder and decoder are connected with residual connections.

In this work, we propose using HuBERT representations at different resolutions (HuBERT-MR) for downstream speech tasks. In our experiments, we evaluate both the parallel and the hierarchical approaches to efficiently utilize HuBERT of different resolutions. Experiments show that our proposed method could get significantly better performances over the original HuBERT at 20ms resolution. In some tasks, the HuBERT with multi-resolution can even achieve reasonable performances compared to large models, even with less training data and fewer parameters.

\section{HuBERT with Multiple Resolutions}

Let $S \in \mathbb{R}^{1 \times L}$ be a speech signal with length $L$. The HuBERT model $H$ consists of two modules: a conv. feature extractor and an $N$-layer transformer encoder. The conv. block first converts $S$ into a sequence of vectors $X^0 = [x^0_1, ..., x^0_T] \in \mathbb{R}^{T \times D}$, where $T$ is the number of frames and $D$ is the dimension of each frame. The resulting feature sequence $X^0$ is then passed to the transformer encoder, which produces a sequence of feature representations $X^i = [x^i_1, ..., x^i_T] \in \mathbb{R}^{T \times D}$ at the $i$-th transformer layer. Each frame $x^{i}_t$ corresponds to a fixed time interval $R$, where $R \cdot T = L$.\footnote{In practical situations, it is necessary to incorporate rounding procedures that take into account edge cases.} We refer to $R$ as the resolution of features.

By controlling the stride of the conv. feature extractor, we can obtain a range of resolutions ($R_1$, ..., $R_K$) and correspondingly, $K$ distinct HuBERT models ($H_1$, ..., $H_K$). In the next subsections, we discuss the application of the parallel and hierarchical approaches discussed in Sec.~\ref{sec: intro} to merge $K$ HuBERT models for downstream tasks. For the easiness of discussion, we consider $K=3$ as an example and all HuBERT models ($H_1$, $H_2$, $H_3$) use the same model configuration for all encoder modules so that the feature dimension for both models is $D$.

\begin{figure}
    \centering
    \includegraphics[width=0.8\linewidth]{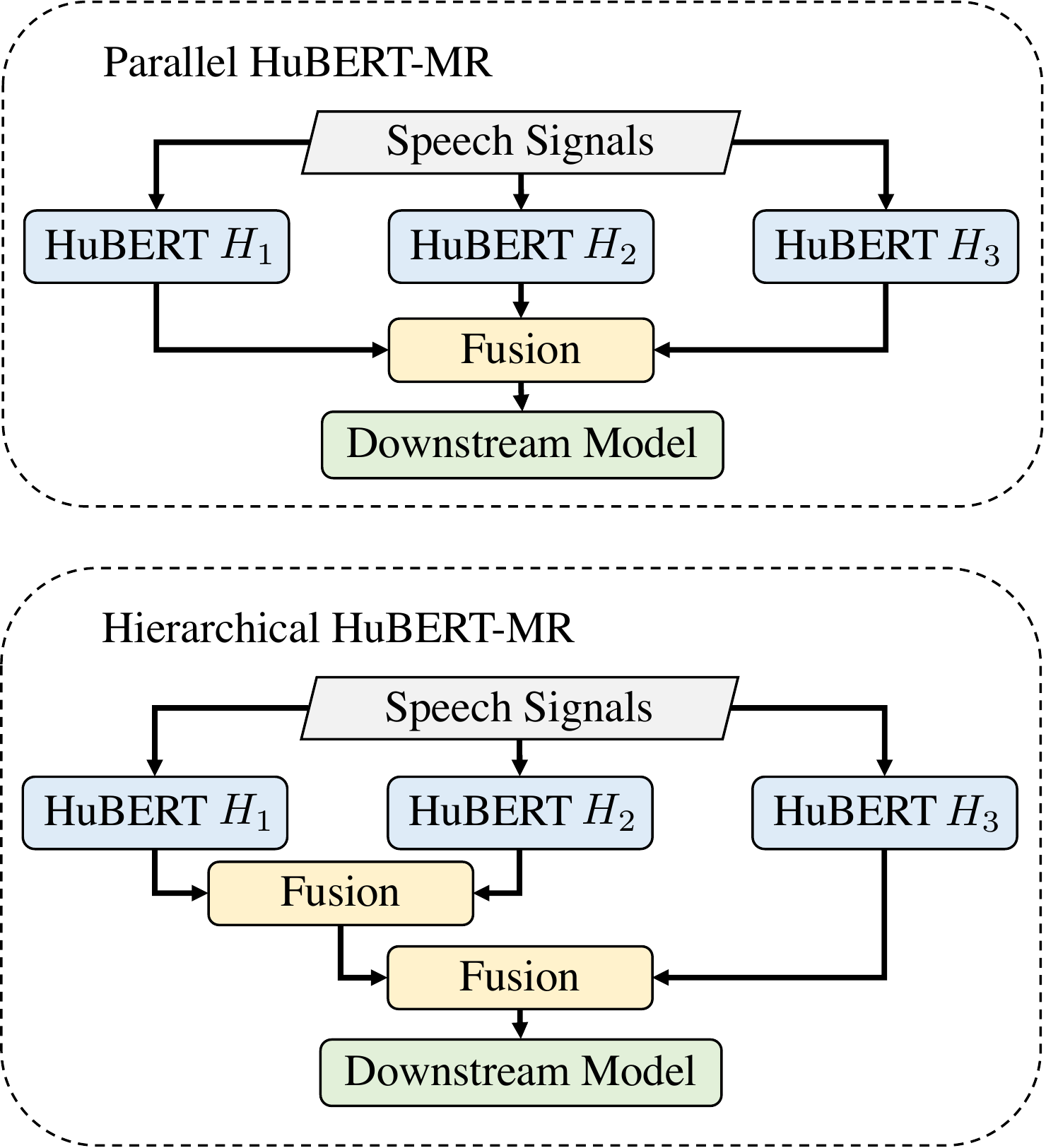}
    \caption{HuBERT-MR-P and HuBERT-MR-H. In HuBERT-MR-P (shown in the upper figure), three HuBERT models are fused in parallel. In contrast, HuBERT-MR-H (shown in the lower figure) fuses HuBERT models hierarchically, with features of low resolutions being fused earlier. Details about the framework design and Fusion modules can be found in Section~\ref{ssec: parallel} and Section~\ref{ssec: hierarchical}, respectively.}
    \label{fig:hubert-mr-hierarchical}
    \vspace{-15pt}
\end{figure}

\subsection{Parallel HuBERT-MR (HuBERT-MR-P)}
\label{ssec: parallel}
As explained in Sec.~\ref{sec: intro}, the parallel approach employs parallel encoders to process input signals at different resolutions. Therefore, we use three HuBERT models ($H_1$, $H_2$, and $H_3$) with resolutions $R_1$, $R_2$, and $R_3$, respectively, to obtain layerwise features from the layer before the top encoder layer to the last encoder layer: $(X_1^0, ..., X_1^N)$, $(X_2^0, ..., X_2^N)$, and $(X_3^0, ..., X_3^N)$. These feature tensors have shapes $X_1^i \in \mathbb{R}^{T_1 \times D}$, $X_2^i \in \mathbb{R}^{T_2 \times D}$, and $X_3^i \in \mathbb{R}^{T_3 \times D}$, respectively. Noted that $T_1, T_2, T_3$ are different. The illustration of HuBERT-MR-P is shown in Figure~\ref{fig:hubert-mr-hierarchical}. We define multi-resolution features $X_{\text{MR-P}}$ as follows:
\begin{equation}
X_{\text{MR-P}} = \sum_{i=0}^N (w_{1,i} \cdot \mathsf{UP}_1 (X_1^i) + w_{2,i} \cdot \mathsf{UP}_2(X_2^i) + w_{3,i} \cdot \mathsf{UP}_3(X_3^i)),
\end{equation}
where $w_{1,i} \in [0, 1]$, $w_{2,i}\in [0, 1]$, and $w_{3,i}\in [0, 1]$ are learnable weights that sum up to one (i.e., $\sum_{i=0}^{N} (w_{1,i} + w_{2,i} + w_{3,i}) = 1$). The functions $\mathsf{UP}_1$, $\mathsf{UP}_2$, and $\mathsf{UP}_3$ upsample the representations to the greatest common divisor of $R_1$, $R_2$, and $R_3$, denoted as $R_{1,2,3}$, to ensure matching feature lengths. Finally, we can use $X_{\text{MR-P}} \in \mathbb{R}^{T_{\text{MR}}\times D}$ for various downstream tasks, where $T_{\text{MR}} = L // R_{1,2,3}$. The $\mathsf{UP}$ functions can be any upsampling functions, including simple methods such as repeating the features along the time axis, as used in \cite{shi2022bridging}, or more complex methods such as transposed conv. networks.


\subsection{Hierarchical HuBERT-MR (HuBERT-MR-H)}
\label{ssec: hierarchical}

As described in Sec.~\ref{sec: intro}, the hierarchical approach models multiple resolutions in a sequential manner. Unlike U-net-based methods \cite{gao2022unet, zhao2021unet++, li2022unet}, we do not perform additional feature encoding as the HuBERT models with different resolutions are already pre-trained. Instead, as shown in Figure~\ref{fig:hubert-mr-hierarchical}, we adopt the decoder architecture inspired by U-net and fuse the HuBERT representations from low to high resolution. Assuming $R_1 > R_2 > R_3$, we first fuse the outputs from $H_1$ and $H_2$, and then we further fuse $H_3$ for additional downstream models.

The fusion module combines the representations from two different resolutions into a single stream. Specifically, we use a conv. module and a de-conv. module for each feature, respectively. Note that additional conv. modules improve the stability of the fusion as observed in our experiments. Given input features $X_1^N \in \mathbb{R}^{T_1 \times D}$ and $X_2^N \in \mathbb{R}^{T_2 \times D}$, we first apply conv. modules with residual connections and then employ transposed conv. modules to align their resolutions. The resulting feature $X_{\text{MR-H}}^{1:2}$ is defined as:
\begin{equation}
    X_{\text{MR-H}}^{1:2} = \text{DeConv}_1(\text{Conv}_1(X_1^N)) + \text{DeConv}_2(\text{Conv}_2(X_2^N)).
\end{equation}
Then, we further apply a conv. module to $X_3^N$ and use transposed conv. modules to compute $X_{\text{MR-H}}^{1:3} \in \mathbb{R}^{T_{\text{MR}} \times D}$ as:
\begin{equation}
    X_{\text{MR-H}}^{1:3} = \text{DeConv}_{1,2}(X_{\text{MR-H}}^{1:2}) + \text{DeConv}_3(\text{Conv}_3(X_3^N)).
\end{equation}
We can then use the feature $X_{\text{MR-H}}^{1:3}$ for downstream tasks.

\section{Experiments}

\subsection{Pre-trained HuBERT}
\begin{table}[t]
    \centering
    \caption{Configurations of the pre-trained HuBERT with multi-resolutions. The convolution (Conv.) module is represented in [(kernel-size, stride)* layer-number].}
    \vspace{-10pt}
    \resizebox {\linewidth} {!} {
\begin{tabular}{l|ccccc}
\toprule
ID & Res.(ms) & Param. & Conv. Module \\
\midrule
\textbf{A} & 20 & 94.7 & (10,5)*1 + (3,2)*4 + (2,2)*2 \\
\textbf{B} & 40 &  95.2 & (10,5)*1 + (3,2)*4 + (2,2)*3 \\
\textbf{C} & 100 & 97.3 & (10,5)*2 + (3,2)*4 + (2,2)*2 \\

\bottomrule
\end{tabular}
}
    \label{tab: config}
    \vspace{-15pt}
\end{table}

\begin{table*}[t]
    \centering
    \caption{HuBERT-MR-P on SUPERB benchmark. Detailed tasks and evaluation metrics are discussed in Sec.~\ref{ssec: exp}. Proposed HuBERT-MR-P is introduced in Sec.~\ref{ssec: parallel}.}
    \vspace{-10pt}
\begin{tabular}{c|cccccccccccc}
\toprule
Model &  Res.(ms) & PR($\downarrow$) & ASR($\downarrow$) & ER($\uparrow$) & IC($\uparrow$) & SID($\uparrow$) & SD($\downarrow$) & SV($\downarrow$) & SE($\uparrow$) & ST($\uparrow$) \\
\midrule
HuBERT & 20 & 5.41 &  6.42 & \textbf{64.92} & 98.34 & 81.42 & 5.88 & 5.11 & \textbf{2.58} & 15.53 \\
wav2vec2 & 20 & 5.74 & 6.43 & 63.43 & 92.35 & 75.18 & 6.08 & 6.02 & 2.55 & 14.81 \\

\midrule
HuBERT-MR-P & (100,40,20)  & \textbf{4.83} & \textbf{5.48} & 63.76 & \textbf{98.51} & \textbf{83.23} & \textbf{5.75} & \textbf{5.10} & 2.55 & \textbf{16.18}\\
\bottomrule
\end{tabular}
    \label{tab: superb}
    \vspace{-15pt}
\end{table*}

To evaluate the effectiveness of HuBERT models with different resolutions, we trained three HuBERT models by modifying their conv. feature extractor. The configurations of these models are presented in Table~\ref{tab: config}. We trained all HuBERT models following the same procedure as HuBERT-base in \cite{hsu2021hubert}, except for changes in label rates and corresponding conv. modules. We conducted two iterations of training for each HuBERT, where the first iteration was trained on Mel frequency cepstral coefficients (MFCC) clusters, and the second iteration was trained using the intermediate features' clusters. The final dimension of each HuBERT was set to $D=768$. We pre-trained all HuBERT models using the Librispeech dataset \cite{panayotov2015librispeech}.

\begin{table}[t]
    \centering
    \caption{The summation of layer weights of HuBERT with different resolutions in the HuBERT-MR-P model, which was evaluated in the SUPERB benchmark as shown in Table~\ref{tab: superb}.}
    \vspace{-10pt}
\begin{tabular}{l|cccc}
\toprule
HuBERT & ASR & SV & ST & Avg. Tasks \\
\midrule
100ms & 0.21 & 0.16 & 0.21 & 0.18 \\
40ms & 0.33 & 0.28 & 0.37 & 0.32 \\
20ms & 0.46 & 0.56 & 0.42 & 0.50 \\
\bottomrule
\end{tabular}

    \label{tab: superb-weight}
    \vspace{-15pt}
\end{table}

\subsection{Experimental Setups}
\label{ssec: exp}

\noindent \textbf{SUPERB benchmark}:
In our experiments, we evaluate HuBERT-MR on the SUPERB benchmark \cite{yang21c_interspeech, tsai2022superb, feng2023superb}. According to the SUPERB benchmark policy, we do not to include additional trainable parameters except for the layerwise weights. Therefore, we mainly focused on HuBERT-MR-P for the SUPERB tasks. We use the repeating method as the upsampling function $\mathsf{UP}$, as described in Sec.~\ref{ssec: parallel}. 

We evaluate most of the SUPERB tasks, including understanding tasks (phone recognition (PR), automatic speech recognition (ASR), intent classification (IC), and speech translation (ST)), speaker-related tasks (speaker identification (SID), speaker verification (SV), and speaker diarization (SD)), and frontend processing (speech enhancement (SE)). 
To better understand the results, we also show the performances on wav2vec2-base for comparison \cite{baevski2020wav2vec, hsu2021hubert}.

\noindent \textbf{ASR Fine-tuning}:
We evaluate the ASR fine-tuning task on the Librispeech 100-hour train set with dev-clean and test-clean for development and testing, respectively. The training utilizes fairseq toolkit \cite{ott2019fairseq}. For all models, we train 100k steps with a maximum token number of 1M to form mini-batches. 
The training uses an AdamW optimizer with 8k warmup steps and a learning rate of 2e-5. We compared HuBERT-MR-P and HuBERT-MR-H with base HuBERT models, as well as wav2vec2 models and HuBERT-large. Instead of repeating, we use transposed convolution for the $\mathsf{UP}$ function of HuBERT-MR-P. For simplification, we use a linear projection for CTC loss computation as the downstream module needed for HuBERT-MR.

We present not only the Viterbi decoding results but also the results after language model rescoring. For decoding, we used Flashlight for wav2letter decoding \cite{kahn2022flashlight} and applied a beam size of 500 with a beam threshold of 100 and a language model weight of 2 for language model rescoring. The language model used was trained on the 4-gram language model training corpus of Librispeech \cite{panayotov2015librispeech}.

\noindent \textbf{Evaluation metrics}
We generally follow the evaluation metrics for SUPERB tasks.
We use phone error rate (PER) for PR; word error rate (WER) for ASR; accuracy for ER, IC, and SID; diarization error rate~(DER) for SD; equal error rate (EER) for SV; PESQ for SE; BLEU for ST. 
While for ASR fine-tuning, we use WER. Meanwhile, for efficiency concerns, we also report Floating Point Operations Per Second (FLOPs) and Multiply-Accumulate Operations (MACs) to models. The calculation procedure follows the SUPERB challenge \cite{feng2023superb}.

\subsection{Experimental Results}

\begin{table}[t]
    \centering
    \caption{Fine-tuning results on Librispeech-100h (comparison with baselines). Results with language model rescoring are in brackets. HuBERT-MR-P is explained in Sec~\ref{ssec: parallel} and HuBERT-MR-H is discussed in Sec~\ref{ssec: hierarchical}.}
    \vspace{-10pt}
\begin{tabular}{c|ccc}
\toprule
Model &  Res.(ms) & WER($\downarrow$) \\
\midrule
HuBERT & 20 & \hphantom{0}7.73(\hphantom{0}3.81) \\
HuBERT & 40 & 12.38(\hphantom{0}4.90) \\
HuBERT & 100 & 98.37(97.87) \\
\midrule
HuBERT-MR-P & (100,20) & \hphantom{0}{6.99}(\hphantom{0}{3.70}) \\
HuBERT-MR-P & (40,20) & \hphantom{0}{7.13}(\hphantom{0}{3.75}) \\
HuBERT-MR-P & (100,40,20) & \hphantom{0}{6.53}(\hphantom{0}{3.61})\\
\midrule
HuBERT-MR-H & (100,20) & \hphantom{0}6.59(\hphantom{0}3.59) \\
HuBERT-MR-H & (40,20) & \hphantom{0}7.01(\hphantom{0}3.71)\\
HuBERT-MR-H & (100,40,20) & \hphantom{0}\textbf{6.11}(\hphantom{0}\textbf{3.31})\\

\bottomrule
\end{tabular}

    \label{tab: librispeech-base}
    \vspace{-15pt}
\end{table}

\begin{table*}[t]
    \centering
    \caption{Fine-tuning results on Librispeech-100h (comparison with other models). Results with language model rescoring are in brackets. * indicates the large model setting. The unlabeled column shows the number of hours used for SSL pre-training. HuBERT-MR-H is discussed in Sec~\ref{ssec: hierarchical}. Noted that the HuBERT base model with 20ms is our baseline.}
    \vspace{-10pt}
\begin{tabular}{c|cc|ccc|c}
\toprule
Model &  Res.(ms) & Unlabeled(h) & Param.(M) & MACs(G) & FLOPs(T) & WER($\downarrow$) \\
\midrule
 HuBERT & 20 & 960 & 94.7 & 1669 & 3.34 & 7.73(3.81) \\
wav2vec2 & 20 & 960 &  95.0 & 1669 & 3.34 & 6.54(4.33) \\
wav2vec2* & 20 & 60K & 317.4 & 4326 & 8.66 & 5.90(3.45)\\
HuBERT* & 20 & 60K & 316.6 & 4324 & 8.66 & \textbf{5.40}(\textbf{2.82}) \\
\midrule
HuBERT-MR-H & (100,40,20) & 960 & 298.4 & 3454 & 6.91  & 6.11(3.31) \\

\bottomrule
\end{tabular}
    \label{tab: librispeech-other}
    \vspace{-15pt}
\end{table*}

\begin{figure*}[ht]
\centering
\subfigure[20ms]{\includegraphics[width=0.3\linewidth]{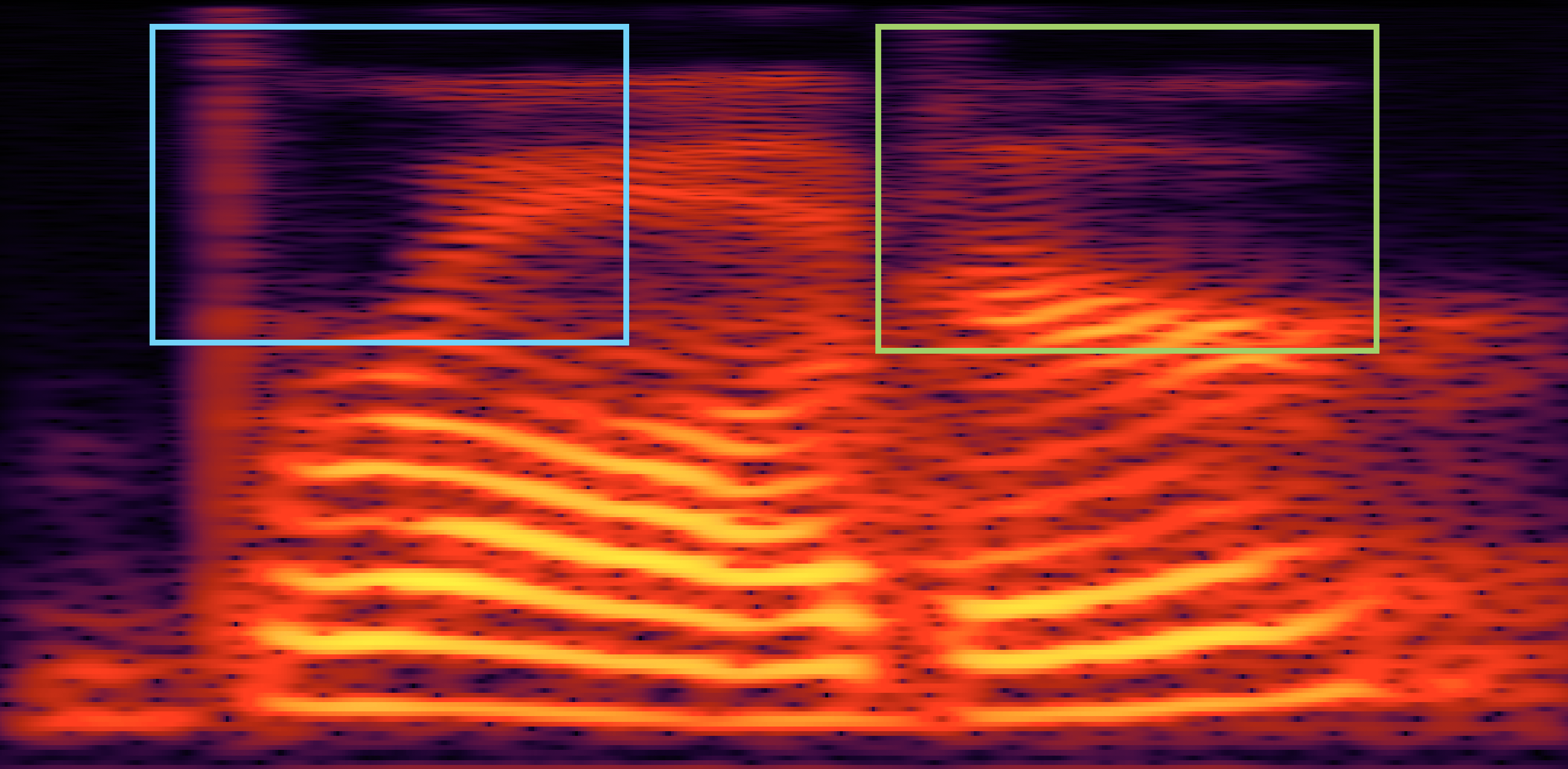}}
\hfill
\subfigure[40ms]{\includegraphics[width=0.3\linewidth]{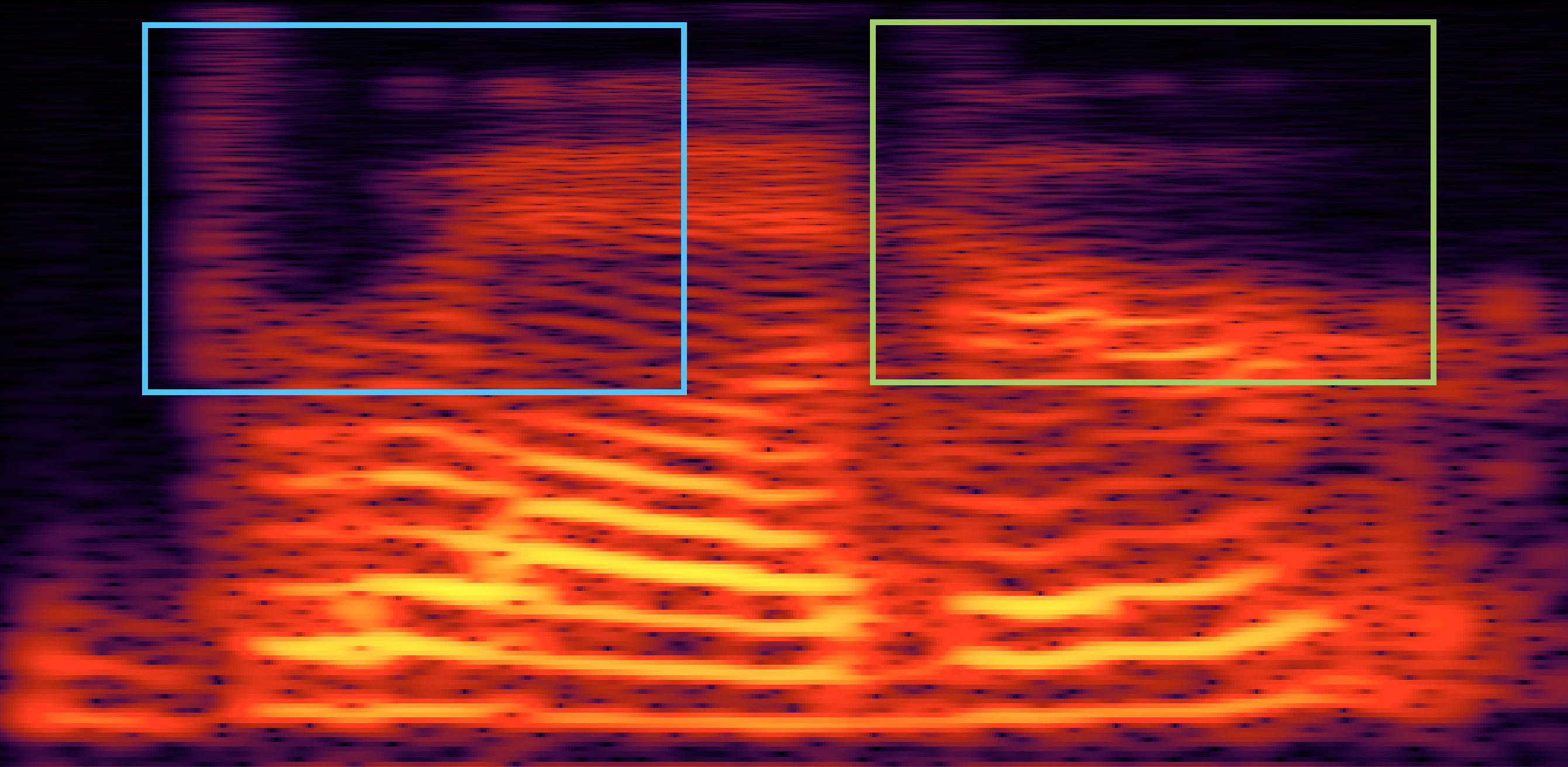}}
\hfill
\subfigure[100ms]{\includegraphics[width=0.3\linewidth]{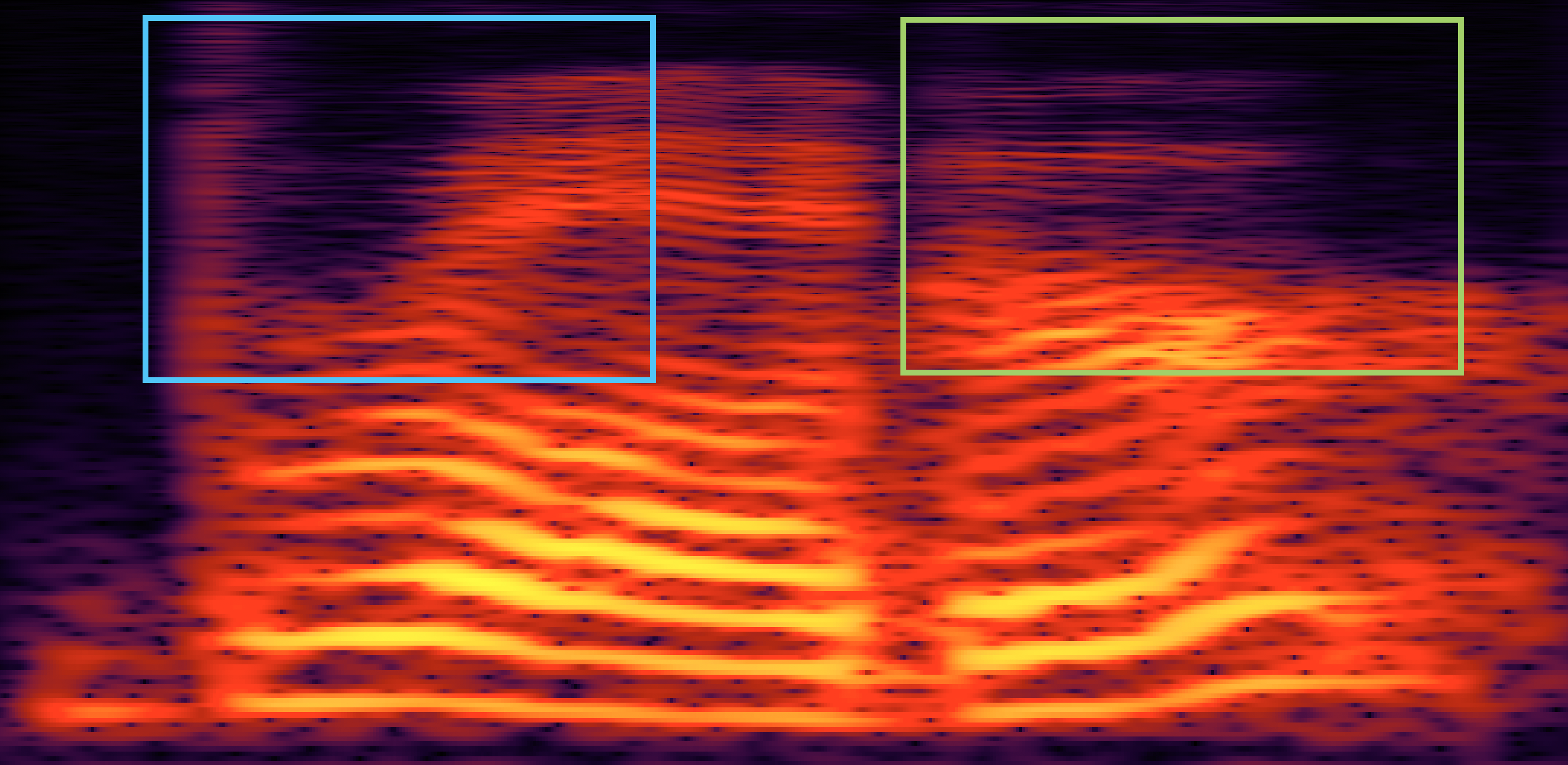}}
\vspace{-10pt}
\caption{Speech re-synthesis using features from HuBERT base models at different resolutions. The high-resolution features capture better envelope information in the time domain (shown in the blue box), while the low-resolution features provide more detailed information in the frequency domain (shown in the green box). See Sec.~\ref{ssec: discussion} for experimental details and discussion.}
\label{fig:threeimages}
\vspace{-15pt}
\end{figure*}

Table~\ref{tab: superb} presents the experimental results of the SUPERB benchmark. In most tasks, HuBERT-MR-P showed significant improvements over the original HuBERT, with the exception of ER and SE. We also analyzed the weight contribution of HuBERT with different representations of HuBERT-MR-P on ASR, SV, ST, and the average overall tasks, which are presented in Table~\ref{tab: superb-weight}. Among the tasks, SE has the lowest weight for 100ms HuBERT (0.15), while ER has the highest weight for 100ms HuBERT (0.26). The results indicate that HuBERT features from multiple resolutions provide additional benefits and can significantly contribute to various types of tasks. 

The experimental results of ASR fine-tuning are presented in Tables \ref{tab: librispeech-base} and \ref{tab: librispeech-other}. Table \ref{tab: librispeech-base} compares the performance of HuBERT-MR with the original HuBERT models. The following observations can be found:
\begin{itemize}
    \item Both HuBERT-MR-P and HuBERT-MR-H outperform the base HuBERT model trained with resolutions of 100ms, 40ms, and 20ms.
    \item Although the HuBERT model trained with resolutions of 100ms and 40ms does not achieve similar performance to the one trained with 20ms, their features appear to be complementary to each other, resulting in improved performance for all HuBERT-MR models. We observe that the 100ms-based HuBERT model does not perform well in the task, likely due to the feature sequence being too short for effective CTC-based training.
\end{itemize}
In Table~\ref{tab: librispeech-other}, we compare the performance of HuBERT-MR-H to that of the HuBERT-large and wav2vec2 models. Our findings are as follows:
\begin{itemize}
    \item HuBERT-MR-H outperforms both the base versions of HuBERT and wav2vec2, highlighting the superior performance of this method.
    \item Although HuBERT-MR-H is a significant improvement over the base HuBERT model, there is still some performance gap when compared to HuBERT-large. This difference could be due to the limited training data (960 Librispeech training sets versus 60K Librilight \cite{kahn2020libri}) and fewer pre-training iterations (all three base HuBERT models use two iterations, while HuBERT-large uses three iterations).
    \item HuBERT-MR-H has a similar parameter size to both HuBERT-large and wav2vec2-large after combining three HuBERT models. However, it requires less computational overhead compared to other large models. This reduction is mainly due to the $O(T^2)$ complexity of the transformer layers in computing intermediate hidden representations \cite{yen2023compress}. While HuBERT-MR-H has lower-resolution networks in its sub-module, it can save computational effort, as shown in MACs and FLOPs in Table~\ref{tab: librispeech-other}.
\end{itemize}

\subsection{Further Discussion}
\label{ssec: discussion}
Our experiments show that HuBERT models with different resolutions can extract features from the same speech source in distinct ways that are useful for downstream tasks. To investigate these differences, we analyzed the features extracted by three pre-trained HuBERT models with 100ms, 40ms, and 20ms resolutions. Specifically, we extracted the 6$^{\text{th}}$ layer representations and used them as input features to train a HiFi-GAN vocoder \cite{kong2020hifi} with the ParallelWaveGAN toolkit \cite{hayashi2020espnet, hayashi2021espnet2}.\footnote{\scriptsize{\url{https://github.com/kan-bayashi/ParallelWaveGAN}}.} We trained the vocoder on the LJSpeech dataset and adapted the upsampling modules to match the resolution of each HuBERT model. The vocoder was trained for 50k steps using the same configuration as the ParallelWaveGAN toolkit. Finally, we generated and compared the spectrograms of synthesized test-set speech produced from different representations, as shown in Figure~\ref{fig:threeimages}.\footnote{Synthesized audio examples can be found in the \scriptsize{\url{https://www.dropbox.com/s/61ap65iegii93il/audio-samples-resynthesis.zip}.} } The followings are some interesting findings:
\begin{itemize}
    \item HuBERT features at different resolutions are capable of producing high-quality re-synthesized speech. Despite not performing well on ASR fine-tuning tasks (as shown in Table~\ref{tab: librispeech-base}), HuBERT with 100ms resolution exhibits excellent speech re-synthesis quality. This suggests that the feature still contains the necessary information in the speech.
    \item As shown in Figure~\ref{fig:threeimages}, high-resolution HuBERT features capture better envelope information in each frame of the speech, while low-resolution features have a more detailed formant presentation. This leads us to hypothesize that high-resolution HuBERT may have a better understanding in the time domain, while low-resolution features have more detailed information in the frequency domain. This property is similar to Short-time Fourier transformation with different window sizes and shifts, to some extent.
\end{itemize}

\section{Conclusion}
In this study, we revisit the use of HuBERT with multiple resolutions, recognizing that the original 20ms resolution may not be optimal for various downstream tasks. To address this, we propose HuBERT-MR, which integrates information from three HuBERT base models pre-trained with different resolutions. We examine two approaches for integration: a parallel approach (HuBERT-MR-P) and a hierarchical approach (HuBERT-MR-H). We evaluate HuBERT-MR-P on the SUPERB benchmark and both HuBERT-MR models on ASR fine-tuning. Our experiments demonstrate that the HuBERT-MR models significantly improve model performance on various downstream tasks, indicating that pre-trained features from multiple resolutions are complementary. Furthermore, we find that HuBERT-MR can outperform larger models in some scenarios, even with less pre-training data and fewer parameters. To further highlight the differences among HuBERT features at different resolutions, we conduct speech re-synthesis with the HiFi-GAN vocoder. Our results demonstrate that the features do differ across resolutions, while all retain the essential information for intelligibility. We believe this work offers valuable insights into the potential benefits of considering multi-resolution SSL in the field.

\section{Acknowledgement}
This work was supported by a Meta AI SRA grant. J. Shi and S. Watanabe are funded in part of the Delta project, supported by the NSF (award OCI 2005572), and the State of Illinois.

\section{References}
{
\printbibliography
}
\end{document}